\documentclass{ieeeaccess}
\usepackage{cite}
\usepackage{amsmath,amssymb,amsfonts}
\usepackage{algorithm}
\usepackage{algpseudocode}
\usepackage{graphicx}
\usepackage{caption}
\usepackage{subcaption}
\usepackage{textcomp}

\usepackage{multirow}
\usepackage{makecell}
\usepackage[group-separator={,}]{siunitx}
\usepackage{hyperref}

\def\BibTeX{{\rm B\kern-.05em{\sc i\kern-.025em b}\kern-.08em
T\kern-.1667em\lower.7ex\hbox{E}\kern-.125emX}}
\begin{document}
\history{Date of publication 26 July 2022, date of current version 3 August 2022.}
\doi{10.1109/ACCESS.2022.3193937}

\title{Accelerating CPU-Based Sparse General Matrix Multiplication With Binary Row Merging}

\author{
\uppercase{Zhaoyang~Du}\authorrefmark{1},~\IEEEmembership{Graduate Student Member, IEEE}
\uppercase{Yijin~Guan}\authorrefmark{2},
\uppercase{Tianchan~Guan}\authorrefmark{2}, 
\uppercase{Dimin~Niu}\authorrefmark{2},
\uppercase{Hongzhong~Zheng}\authorrefmark{2},
\uppercase{Yuan~Xie}\authorrefmark{2},~\IEEEmembership{Fellow, IEEE}
} 
\address[1]{College of  Information  Science  and  Electronic Engineering, Zhejiang University, Hangzhou 310007, China}
\address[2]{Alibaba Group, Hangzhou 311121, China}


\markboth
{This paper has been accepted by IEEE Access (DOI:10.1109/ACCESS.2022.3193937).}
{This paper has been accepted by IEEE Access (DOI:10.1109/ACCESS.2022.3193937).}

\corresp{Corresponding author: Zhaoyang Du (e-mail:11731021@zju.edu.cn).}

\begin{abstract}
Sparse general matrix multiplication (SpGEMM) is a fundamental building block for many real-world applications. Since SpGEMM is a well-known memory-bounded application with vast and irregular memory accesses, considering the memory access efficiency is of critical importance for SpGEMM's performance. Yet, the existing methods put less consideration into the memory subsystem and achieved suboptimal performance. In this paper, we thoroughly analyze the memory access patterns of SpGEMM and their influences on the memory subsystem. Based on the analysis, we propose a novel and more efficient accumulation method named BRMerge for the multi-core CPU architectures.

The BRMerge accumulation method follows the row-wise dataflow. It first accesses the $B$ matrix, generates the intermediate lists for one output row, and stores these intermediate lists in a consecutive memory space, which is implemented by a ping-pong buffer. It then immediately merges these intermediate lists generated in the previous phase two by two in a tree-like hierarchy between two ping-pong buffers. The architectural benefits of BRMerge are 1) streaming access patterns, 2) minimized TLB cache miss rate, and 3) reasonably high L1/L2 cache hit rates, which result in both low access latency and high bandwidth utilization when performing SpGEMM. Based on the BRMerge accumulation method, we propose two SpGEMM libraries named BRMerge-Upper and BRMerge-Precise, which use different allocation methods. Performance evaluations with 26 commonly used benchmarks on two CPU servers show that the proposed SpGEMM libraries significantly outperform the state-of-the-art SpGEMM libraries.

\end{abstract}


\titlepgskip=-15pt

\maketitle

\section{Introduction}\label{sec:intro}
\PARstart{S}{parse} general matrix multiplication (SpGEMM) is a fundamental building block in many real-world applications such as algebraic multigrid solvers~\cite{AMG, AMG2}, multi-source breadth first search~\cite{BFS}, Markov clustering~\cite{markov}, finite element simulations based on domain decomposition~\cite{finite-element}, and molecular dynamics simulations~\cite{molecular}. 

The extensive use of SpGEMM in real-world applications has led to the development of several SpGEMM libraries on CPUs~\cite{gustavson, matlab, yusuke, PB-SpGEMM}, GPUs~\cite{cusp, bhsparse, rmerge, rmerge2, nsparse, speck}, and accelerators~\cite{outerspace, sparch, matraptor, gamma}, targeting at high-performance computing. In this paper, we focus on optimizing SpGEMM libraries on multi-core CPU architectures.

As a memory-bounded application, the memory access efficiency of both the main memory and multi-level caches is critical for SpGEMM's performance~\cite{PB-SpGEMM, roofline}. 
There are two critical memory access patterns in performing SpGEMM. The first pattern is the vast and irregular memory accesses to the $B$ matrix for the row-wise SpGEMM~\cite{yusuke}, or the $\hat{C}$ matrix for the outer-product SpGEMM~\cite{PB-SpGEMM}. These memory accesses usually span large memory spaces, which is not friendly for the memory system (especially the translation lookaside buffer (TLB)~\cite{intel-opt}) of modern multi-core CPU architectures.
The second pattern is the memory accesses when accumulating multiple intermediate lists with varying lengths into one result row. These intermediate lists should be accessed multiple times with irregular access patterns due to the inherent sorting property of the accumulating operation. Hence, the second memory access pattern is also not friendly for the memory system (especially the L1/L2 caches) of modern multi-core CPU architectures.

The existing Heap-SpGEMM~\cite{yusuke} and Hash-SpGEMM~\cite{yusuke} only consider the computation complexity while putting no consideration to the memory access efficiency mentioned above. Specifically, the Heap-SpGEMM may have high TLB cache miss rate when accessing the $B$ matrix, whereas the Hash-SpGEMM may waste memory bandwidth utilization when accumulating the intermediate lists due to the random access pattern caused by hashing operations. As a result, the two libraries suffers suboptimal performance on the CPU architectures (Section~\ref{sec:result}). 

Although the PB-SpGEMM~\cite{PB-SpGEMM} considers the memory access efficiency, they fail to fully assess the gains and overheads in their method. They focus on solving the vast and irregular memory accesses to the $B$ matrix in the row-wise dataflow by adopting the outer-product dataflow~\cite{yusuke, PB-SpGEMM}. However, their memory accesses to the $\hat{C}$ matrix are still irregular with an even larger amount~\cite{PB-SpGEMM}. Therefore, the PB-SpGEMM may suffer worse TLB cache hit rate than the Heap-SpGEMM. As a result, PB-SpGEMM suffers exceptionally lower performance than its precedent Heap-SpGEMM and Hash-SpGEMM (Section~\ref{sec:result}).

To avoid or minimize the inefficient memory access issues in the existing SpGEMM libraries, we propose a novel and more efficient accumulation method, named BRMerge, for SpGEMM on the CPU architectures. To better understand the proposed work and its architectural benefits, we first describe several backgrounds and the BRMerge accumulation method. And then, we discuss the architectural benefits of the BRMerge accumulation method and the inefficient memory access issues of the existing methods in Section~\ref{sec:memory}. We denote the proposed accumulation method as BRMerge since it is based on the binary-row-merging algorithm described in Section~\ref{sec:brmerge}.

The BRMerge accumulation method adopts the row-wise dataflow to perform SpGEMM, which multiplies each row of $A$ with the whole $B$ matrix for the corresponding row of $C$~\cite{gustavson, yusuke, nsparse, speck}. It consists of a multiplying phase and an accumulating phase to compute each output row. In the first multiplying phase, the multiple required rows of $B$ are accessed, multiplied with the nonzeros in the corresponding row of $A$, and the generated intermediate lists are stored in a consecutive memory space, which is implemented with a ping-ping buffer. In the second accumulating phase, the multiple intermediate lists generated in the previous phase are merged two by two in a tree-like hierarchy between two ping-pong buffers. 
The architectural benefits of BRMerge are the streaming access patterns, minimized TLB misses, and reasonably high L1/L2 cache hit rates, which result in both low access latency and high bandwidth utilization when performing SpGEMM on the multi-core CPU architectures (Section~\ref{sec:memory}).

Based on the BRMerge accumulation method, we further propose two SpGEMM libraries named BRMerge-Upper and BRMerge-Precise. The BRMerge-Upper and BRMerge-Precise use the upper-bound~\cite{bhsparse} and precise allocation methods~\cite{nsparse}, respectively. Comprehensive evaluations with 26 commonly used benchmarks on two Intel Xeon CPUs show that the proposed SpGEMM libraries outperform the state-of-the-art SpGEMM libraries significantly. Specifically, BRMerge-Precise achieves on average $1.42\times$ and $1.39\times$ performance speedups compared to the state-of-the-art best-performing SpGEMM library (i.e., Hash-SpGEMM~\cite{yusuke}) on the Intel Xeon Platinum 8163 CPU and the Intel Xeon Gold 6254 CPU, respectively.

The key technical contributions of this work are as follows:
\begin{itemize}
\item We present a novel accumulation method, named BRMerge, for the SpGEMM algorithm on multi-core CPU architectures.
\item We analyze the architectural benefits of the proposed BRMerge accumulation method on multi-core CPU architectures, emphasizing the memory subsystem. 
\item We propose two SpGEMM libraries named BRMerge-Upper and BRMerge-Precise based on the BRMerge accumulation method and different allocation methods.
\item Evaluations with 26 commonly used benchmarks on two Intel Xeon CPUs show that the proposed SpGEMM libraries outperform the state-of-the-art SpGEMM libraries significantly.
\end{itemize}

The rest of this paper is organized as follows. Section~\ref{sec:bg}  provides backgrounds and related work on the SpGEMM algorithm. Section~\ref{sec:method} describes the binary-row-merging algorithms, the proposed BRMerge accumulation method, the performance analysis of the BRMerge accumulation method, and two proposed SpGEMM libraries based on the BRMerge accumulation method and other methods. Section~\ref{sec:evaluate} first describes the performance evaluation environments, then provides the results and discussions. Section~\ref{sec:conclude} concludes this paper.

The source code of this paper is provided in \url{https://github.com/lorentzbf/BRMerge.git}

\section{Backgrounds and Related Work}\label{sec:bg}
In this section, we provide the backgrounds and related work of SpGEMM algorithms mainly on CPUs.
\subsection{S\lowercase{p}GEMM Preliminaries}
\subsubsection{Notations}\label{sec:note}
We define several notations for the rest of this paper. Capital letters $A$, $B$, and $C$ denote matrices. The two input matrices are $A$ and $B$ with sizes of $M \times K$ and $K \times N$, respectively. The result matrix is denoted as $C$ with a size of $M \times N$.

$A_{ij}$ represents an element located at the $i_{th}$ row and $j_{th}$ column of $A$. $A_{i*}$ represents all the nonzero elements in the $i_{th}$ row of $A$. Similarly, $B_{*j}$ represents all the nonzero elements in the $j_{th}$ column of $B$.

\subsubsection{Dataflows to Perform S\lowercase{p}GEMM}\label{sec:dataflow}
Given two input matrices $A$ and $B$, the text-book definition of matrix multiplication is computed as:
\begin{equation}C_{ij} = \sum_{k= 1}^{K}A_{ik}\cdot{B}_{kj}.\label{eq:inner}\end{equation}
Equation~(\ref{eq:inner}) describes the inner-product dataflow to perform SpGEMM.

One variation of Equation~(\ref{eq:inner}) is:
\begin{equation}C_{i*} = \sum_{k}A_{ik}\cdot{B}_{k*},\label{eq:row-wise}\end{equation}
where $k$ belongs to the set of column indices of the nonzero elements in each row of $A$.
Equation~(\ref{eq:row-wise}) describes the row-wise dataflow to perform SpGEMM. A similar variation of the row-wise dataflow is the column-wise dataflow described in Equation~(\ref{eq:column-wise}):
\begin{equation}C_{*j} = \sum_{k}A_{*k}\cdot{B}_{kj},\label{eq:column-wise}\end{equation}
where $k$ is the set of row indices in each column of $B$.

The SpGEMM can also be computed as:
\begin{equation}C = \sum_{k = 1}^{K}A_{*k}\cdot{B}_{k*},\label{eq:outer}\end{equation}
where the $\sum$ operation computes the sum of multiple sparse partial matrices. Equation~(\ref{eq:outer}) describes the outer-product dataflow to perform SpGEMM.

All the four dataflows mentioned above can generate the correct result matrix $C$ but with different performance issues on modern hardware platforms. Generally, the row-wise dataflow is reported to have the best performance than other dataflows by recent SpGEMM work on CPUs~\cite{yusuke}, GPUs~\cite{rmerge, rmerge2, nsparse, speck} and accelerators~\cite{matraptor, gamma}. Note that, based on Equation~(\ref{eq:row-wise}) and Equation~(\ref{eq:column-wise}), the implementation of the row-wise and column-wise dataflows are dual of each other. Therefore, we only discuss the row-wise dataflow for simplicity. 

There are three key benefits of the row-wise dataflow compared to the inner-product dataflow: 1) zero elements are entirely skipped in both the computation and memory access, 2) the computation of each output row is independent of each other; therefore, the row-wise SpGEMM can be easily parallelized, and 3) accumulating the intermediate lists of each output row has a good temporal locality. As a result, we adopt the row-wise dataflow in our proposed SpGEMM algorithm.

\subsubsection{CSR Storage Format}\label{sec:CSR}
The compressed sparse row (CSR) storage format is one of the most commonly used sparse storage formats for SpGEMM, which is commonly used in the row-wise SpGEMM~\cite{yusuke, PB-SpGEMM, bhsparse, rmerge, rmerge2, nsparse, speck}.

\begin{figure}[h]
\centering
\includegraphics[width=0.45\textwidth]{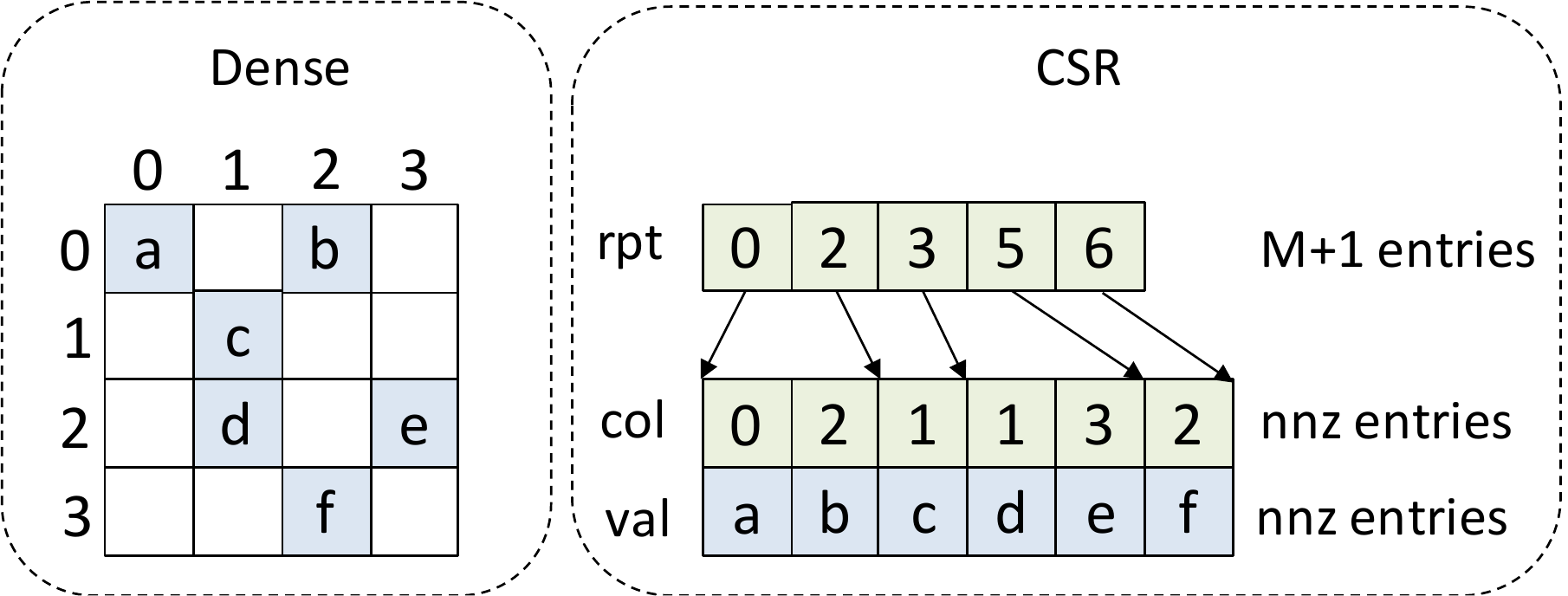}
\caption{Illustration of the CSR storage format. Left: matrix $A$ in the dense storage format. Right: matrix $A$ in the CSR storage format}
\label{fig:CSR}
\end{figure}

Fig.~\ref{fig:CSR} illustrates the CSR storage format. The CSR consists of three arrays to record the nonzero elements and their corresponding indices. The $val$ and $col$ arrays record the values of the nonzero elements and their corresponding column indices in a sorted row-major and column-major order. The lengths of $val$ and $col$ arrays are both the number of nonzero elements ($n_{nz}$) of the sparse matrix. The $rpt$ array records the start and end offsets for each row's values and column indices in the $val$ and $col$ arrays. Since the $i_{th}$ row's end offset can be encoded the same as the $(i+1)_{th}$ row's start offset in the $rpt$ array, the $rpt$ array is further compressed to $M + 1$ entries, where $M$ is the number of rows. One of the key performance benefits of using CSR is that it is easy to access the elements of an entire row. In this paper, we also use CSR as the storage format for the input and output matrices in the proposed SpGEMM libraries.

\subsubsection{Compression Ratio (CR)}\label{sec:CR}
In performing SpGEMM, usually, multiple intermediate products are combined to generate one result element due to the duplicate indices. The compression ratio is defined by dividing the total number of intermediate products ($n_{prod}$) in performing SpGEMM by the result matrix's total number of nonzero elements ($n_{nz}$) (Equation~(\ref{eq:CR})). In other words, the compression ratio shows the average number of intermediate products that generate one nonzero element in the result matrix. 
\begin{equation}Compression\ ratio = \frac{Total\ n_{prod}\ to\ compute\ C}{Total\ n_{nz}\ of\ C}.\label{eq:CR}\end{equation}

\subsection{Related S\lowercase{p}GEMM Designs}
In this section we describe the existing methods to address two important design issues: the accumulation method and the allocation method. The accumulation method generates multiple intermediate lists and merges these lists to one result row. Since the structure of the result matrix cannot be known in advance, where to store the result row in the accumulation method remains a challenging task. The allocation method tackles the unknown memory allocation challenge mentioned above.
\subsubsection{Accumulation Methods}\label{sec:accumulation}

The heap-based accumulation method~\cite{yusuke} merges all the intermediate lists of one result row simultaneously by using a heap data structure. Since the intermediate lists are already in sorted ascending order, the heap keeps all the intermediate lists' front elements to sort the result row. In each iteration, the heap pops the smallest element to the result row. The popped element will be added to the current position in the result row or be stored in a new position in the result row, based on the indices of the popped element and the current element in the result row. After this operation, the heap reads the next element from the intermediate list which provides the smallest element last time. The read element should be pushed to the heap for the next iteration.

The most two time-consuming operations in the heap-based accumulation method are the pop and push operations, which are both with $O(log(k))$ complexity, where $k$ is the heap size. In this paper, we use the binary-row-merging-based accumulation method, which can be seen as a heap-based method when the heap size is two. As a result, our accumulation method's pop and push operations are simply one comparison operation and one pointer addition operation for the two input intermediate lists. 

The hash-based accumulation method~\cite{yusuke} merges the intermediate lists by inserting all the intermediate elements into a hash table. After the insertion, the valid nonzero elements are extracted and sorted to generate one result row~\cite{yusuke, nsparse}. The hashvec-based accumulation method~\cite{yusuke} follows the same computation flow as the hash-based accumulation method. The difference is that the hashvec-based accumulation method uses the Intel's SIMD instructions~\cite{wiki-AVX} to implement the hashing operation. The hashvec-based accumulation method is reported to have better performance than the hash-based counterpart when the hash collision rate is high~\cite{yusuke}.

The ESC-based accumulation method is first proposed in the GPU work~\cite{cusp} and is recently used by the CPU work~\cite{PB-SpGEMM} with the outer-product dataflow. The ESC represents \textbf{E}xpand, \textbf{S}ort, and \textbf{C}ompress. It first expands the intermediate lists by storing them consecutively without order. Then it sorts the unordered intermediate lists with duplicate column indices. At last, it compresses the sorted list by merging the duplicate elements which are consecutive to each other.

The row-merge-based accumulation method is proposed on the GPU architectures~\cite{rmerge, rmerge2}. It merges a pre-defined power-of-two number (e.g., 2, 4, 8, 16 et al.) of intermediate lists simultaneously using subwarp~\cite{cuda-doc}. It is similar to the heap-based accumulation method in that it merges multiple intermediate lists simultaneously. Since it uses one thread to maintain one or more intermediate lists and accesses each of these intermediate lists consecutively within the thread, the warp-level accesses to these intermediate lists are completely non-coalesced, which is detrimental to its performance on GPU architectures.

\subsubsection{Allocation Methods}\label{sec:allocation}
The structure of the result matrix (i.e., the number of nonzero elements per output row) cannot be determined before performing SpGEMM. Hence, allocating the memory space to store the result matrix, which requires information of the output structure, is a challenging task. We describe two commonly used allocation methods to tackle this challenge. The two allocation methods are also implemented in the two proposed SpGEMM libraries.

The upper-bound allocation method~\cite{bhsparse, yusuke} first computes the number of intermediate products ($n_{prod}$) per output row as the upper-bound output structure, which is used for the memory allocation for the result matrix in an temporary memory space. The computation of the $n_{prod}$ is a very low-cost operation compared to SpGEMM. And then, the result matrix is computed and stored in the temporary memory space. Since the result matrix should be stored in a standard storage format (e.g., CSR format), an additional memory copy operation is required to move the result matrix from the temporary memory space to the memory space conforming to the standard storage format. 

The precise allocation method~\cite{yusuke, nsparse, speck} first computes the number of nonzero elements ($n_{nz}$) of each result row. This process only computes the indices information of the input matrices without floating-point multiplication, which is commonly named as the symbolic computation~\cite{nsparse}. Then the precise memory space can be allocated based on a low-cost prefix-sum operation on the computed $n_{nz}$ information. At last, the result matrix is computed and directly stored in the standard memory space without the need for the additional copy operation. However, one additional cost when using the precise allocation method is the nontrivial symbolic computation, which has a similar complexity to the SpGEMM itself.

\section{Method}\label{sec:method}
In this section, we first describe the binary-row-merging algorithm and the BRMerge accumulation method. And then we analyze the architectural benefits of BRMerge and the inefficient memory access issues of the existing methods. At last, we describe two proposed SpGEMM libraries, named BRMerge-Upper and BRMerge-Precise, based on the BRMerge accumulation method.

\subsection{Binary-row-merging Algorithm}\label{sec:brmerge}
\begin{figure}[h]
\centering
\includegraphics[width=0.45\textwidth]{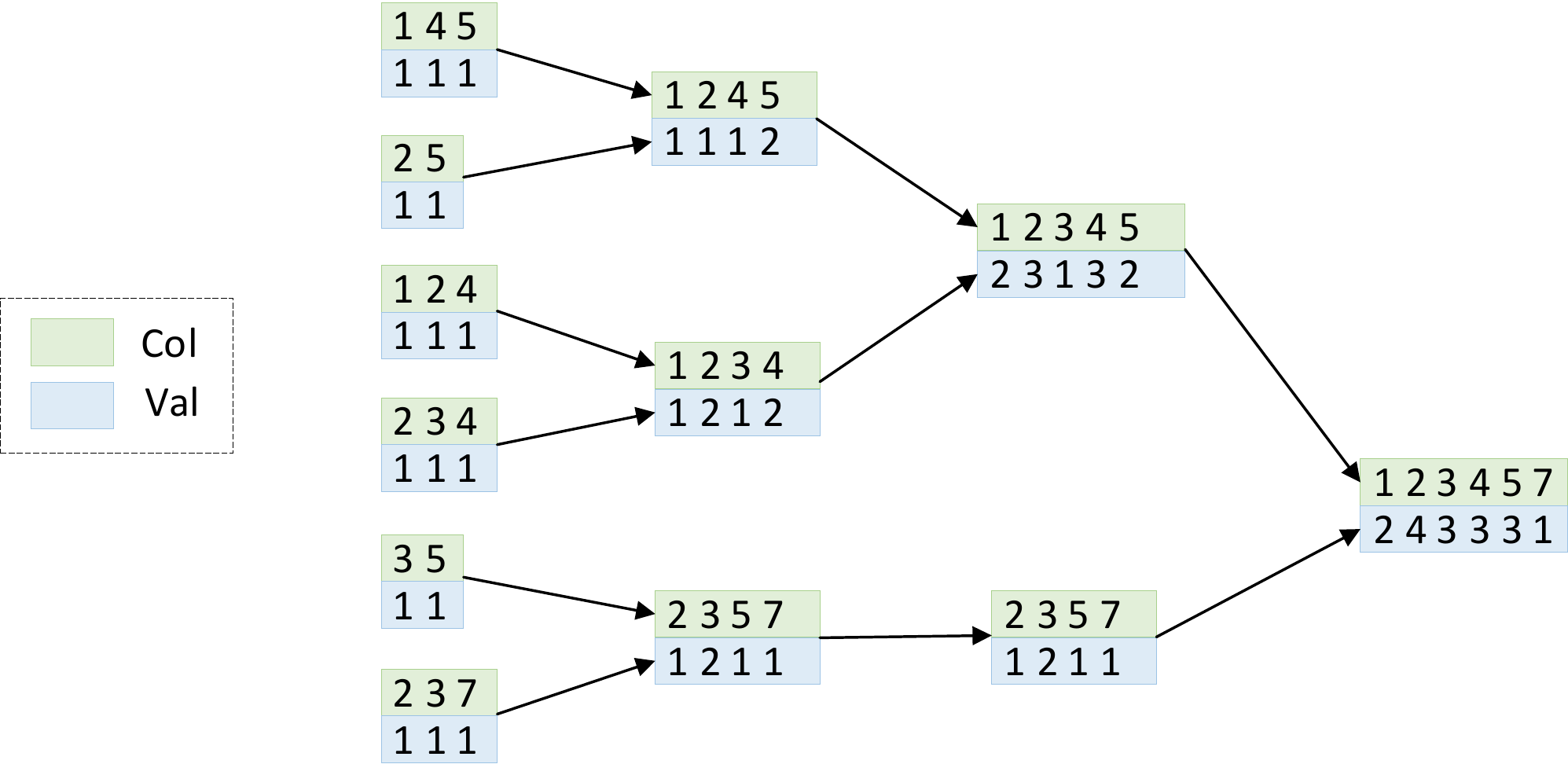}
\caption{Illustration of the binary-row-merging algorithm. There are six sorted lists. the elements of the lists are $col$ and $val$ pairs, where $col$ represents the column indices and $val$ represent the values. Two lists are merged each time by comparing the $col$s of two front elements, which is similar to the mergesort algorithm~\cite{merge-sort}, except that the $val$s with duplicate $col$s are combined into one $val$. All the lists are merged two by two in a tree-like hierarchy. After each round of merging, the number of intermediate lists is reduced by half. Therefore, there are totally $\lceil log(6)\rceil = 3$ rounds of merging until one list is left as the result list.}
\label{fig:brmerge}
\end{figure}

The binary-row-merging algorithm merges multiple sorted lists into one list. Fig.~\ref{fig:brmerge} illustrates how the binary-row-merging algorithm merges six sorted lists into one list. 

\subsection{BRM\lowercase{erge} Accumulation Method}\label{sec:brmerge-accumu}
Algorithm~\ref{alg:brmerge} illustrates the proposed BRMerge accumulation method in a single thread version. BRMerge mainly consists of two phases: 1) generating the intermediate lists, named multiplying phase, and 2) merging the generated lists into one result row, named accumulating phase. 

\begin{algorithm}
\caption{BRMerge accumulation method}
\label{alg:brmerge}
\begin{algorithmic}[1]
	\Require A, B in CSR format
	\State {Compute max\_row\_nprod\_C, max\_row\_nnz\_A}
	\State {ping\_buffer = \textbf{new double} [max\_row\_nprod\_C]}
	\State {pong\_buffer = \textbf{new double} [max\_row\_nprod\_C]}
	\State {ping\_list\_offset = \textbf{new int} [max\_row\_nnz\_A + 1]}
	\State {pong\_list\_offset = \textbf{new int} [max\_row\_nnz\_A + 1]}
	\For {$i \gets 1..M$ } 
	\State{dst\_buffer = ping\_buffer}
	\State{dst\_list\_offset = ping\_list\_offset}
	\State{buffer\_incr = 0, list\_incr = 0, dst\_list\_offset[0] = 0}
	\For {$A_{ik}$ in $A_{i*}$} \Comment{Multiplying phase}
	\For{$B_{kj}$ in $B_{k*}$}
	\State {dst\_buffer[buffer\_incr++] = $A_{ik}$ * $B_{kj}$}
	\EndFor
	\State {dst\_list\_offset[++list\_incr] = buffer\_incr}
	\EndFor
	\State {num\_list = list\_incr}
	\State {src\_buffer = dst\_buffer}
	\State {src\_list\_offset = dst\_list\_offset}
	\State {dst\_buffer = pong\_buffer}
	\State {dst\_list\_offset = pong\_list\_offset}
	\While {num\_list $>$ 1} \Comment{Accumulating phase}
	\State {inner\_num\_list = num\_list, num\_list = 0}
	\While {inner\_num\_list}
	\If {inner\_num\_list $\geq$ 2}
	\State {consume two lists and store to dst\_buffer}
	\State {inner\_num\_list -= 2}
	\ElsIf {inner\_num\_list == 1}
	\State {copy the last list to dst\_buffer}
	\State {inner\_num\_list -= 1}
	\EndIf
	\State {num\_list++}
	\EndWhile
	\State {swap(src\_buffer, dst\_buffer)}
	\State {swap(src\_list\_offset, dst\_list\_offset)}
	\EndWhile
	\State {Store the $i_{th}$ result row in the src\_buffer into the appropriate memory space}
	\State {Release the allocated ping-pong buffers}
	\EndFor
\end{algorithmic}
\end{algorithm}

Since the accumulating phase in BRMerge consists of multiple rounds of merging, we use ping-pong buffers to store the input and output lists in each round of merging. For simplicity in Algorithm~\ref{alg:brmerge}, the \emph{ping\_buffer} and the \emph{pong\_buffer} represent both the $col$'s and $val$'s pair of the ping-pong buffers. We use an additional pair of ping-pong buffers named \emph{ping\_list\_offset} and \emph{pong\_list\_offset} to record the list offsets of the intermediate lists which are stored consecutively in the \emph{ping\_buffer} and \emph{pong\_buffer}. 

Line 1 in Algorithm~\ref{alg:brmerge} computes the sizes of the two pairs of ping-pong buffers, which are the maximum $n_{prod}$ of the output rows and the maximum $n_{nz}$ of the rows in $A$. Line 2--5 allocates the ping-pong buffers. The multiplying phase (Line 10--15) follows the row-wise dataflow (Equation~(\ref{eq:row-wise})). It multiplies the nonzero elements in each row of $A$ with the corresponding rows in $B$ to generate the intermediate lists. The multiple generated intermediate lists are stored consecutively in one of the ping-pong buffers (pointed by the \emph{dst\_buffer}). The start offset of each list is recorded in one of the \emph{list\_offset} ping-pong buffers  (pointed by the \emph{dst\_list\_offset}). Note that each generated list is naturally sorted since each row of $B$ is sorted due to the CSR format. 

\begin{figure}[h]
\centering
\includegraphics[width=0.45\textwidth]{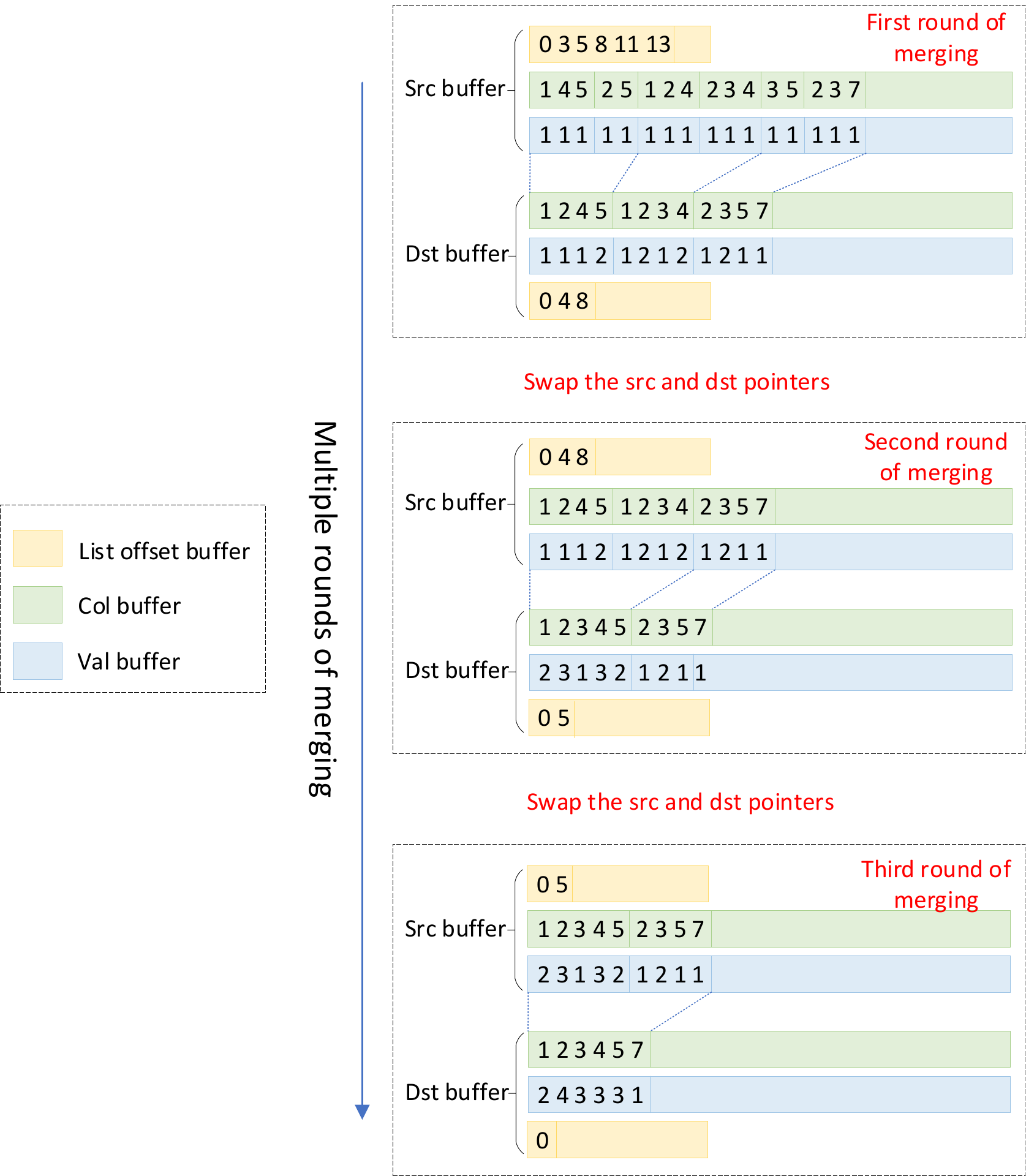}
\caption{Illustration of Line 21--35 in Algorithm~\ref{alg:brmerge}, which is the implementation of the binary-row-merging algorithm. The illustrated implementation uses the same input data as used in Fig.~\ref{fig:brmerge}.}
\label{fig:brmerge-imp}
\end{figure}

The accumulating phase (Line 21--35) implements the binary-row-merging algorithm, which is further illustrated in Fig.~\ref{fig:brmerge-imp}. The accumulating phase is implemented with two while-loops. The outer while-loop determines how many rounds of merging are required. It monitors the \emph{num\_list} information, which is reduced by half after each round of merging. The exit condition of the outer while-loop is when \emph{num\_list} equals $1$ (Line 21). The inner while-loop consumes the \emph{num\_list} intermediate lists from the \emph{src\_buffer} two by two and stores the merged result list to the \emph{dst\_buffer} (Line 25). The consume operation of two input lists is similar to the merge sort algorithm~\cite{merge-sort}, except that the elements with duplicate indices are added together. When only one intermediate list is left in the inner while-loop, it is directly copied to the \emph{dst\_buffer} (Line 28). After the completion of each round of merging (i.e., after the completion of one inner while-loop), the pointers of the ping-pong buffers are swapped (Line 33--34) without data movement. As a result, in each inner while-loop, we always read input lists from the \emph{src\_buffer} and write the result lists to the \emph{dst\_buffer}. After the two while-loops, the result row is stored in the \emph{src\_buffer} due to the last swap operation. Line 36 writes the result row in the \emph{src\_buffer} to the appropriate memory space determined by the allocation method.

\subsection{Performance Analysis of the BRM\lowercase{erge} Accumulation Method and Other Methods}\label{sec:memory}
Two of the most critical memory access patterns in performing SpGEMM are 1) the vast and irregular memory accesses to the $B$ matrix for the row-wise SpGEMM~\cite{yusuke}, or to the $\hat{C}$ matrix for the outer-product SpGEMM~\cite{PB-SpGEMM}, and 2) the memory accesses in the accumulating phase, which accumulates multiple intermediate lists with varying lengths.
In this section, we analyze the architectural benefits of the BRMerge accumulation method for the two critical memory access patterns on modern multi-core CPU architectures. We also analyze the inefficient memory access issues of the existing methods compared to BRMerge.

In the following, we present the analysis mentioned above when we show the three architectural benefits of the proposed BRMerge, which are the streaming access patterns, minimized TLB cache miss rate, and reasonably high L1/L2 cache hit rates.

\subsubsection{Streaming Access Patterns}
The streaming access patterns is mainly in the accumulating phase in BRMerge. The accumulating phase uses one CPU thread to merge all the intermediate lists two by two. Therefore, at any given time, it only merges two lists into one result list. The memory read and write patterns are always consecutive when accessing the input and output lists. Furthermore, the duration of the read and write operation is short since it only performs one simple comparison with one potential addition operation before the next data read and write operations. Therefore, the memory access pattern of the input and output lists is a typical streaming access pattern. 

There are mainly two benefits of the streaming access patterns in modern CPU architectures. The first benefit is to maximize the memory bandwidth utilization between any two high-level memory systems (e.g., between L1 and L2 caches, or between L2 and L3 caches). In modern CPUs, the memory access granularity between the CPU and the L1 cache can be as small as 4 or 8 bytes. However, the minimum memory access granularity between high-level memory systems is usually at least a cache line (e.g., 64 bytes).
When a program randomly accesses a 4-bytes memory space, which loads a cache line from the L2 to the L1, the other 60 bytes of the loaded cache line may not be used after a long duration. And after the long duration, the above-loaded cache line may have been evicted from the L1 cache, which means the bandwidth utilization between the L1 and L2 caches is only $4/64 = 6.25\%$. In contrast, when a program has a streaming access pattern and accesses a 4-bytes memory space, which loads a cache line from the L2 to the L1, the other 60 bytes of the loaded cache line will be used immediately. Therefore, a program with the streaming access patterns can achieve up to $64/64 = 100\%$ bandwidth utilization between the high-level memory systems (e.g., between L1 and L2 caches).

The second benefit of the streaming access pattern is to take advantage of the hardware pre-fetching mechanism, which can increase the L1 cache hit rate. Modern CPUs usually have a hardware pre-fetching mechanism, which automatically pre-fetches the following cache line when a few consecutive memory accesses are detected~\cite{intel-opt}. Consider a typical case where a CPU program consecutively accesses and processes data in a memory space with a size between the L1 and L2 cache sizes. Due to the pre-fetching mechanism, most to-be-accessed data may have been pre-fetched from L2 to L1, improving the L1 cache hit rate. This benefit is provided that the CPU also takes time to process the accessed data; therefore, part or all of the pre-fetching latency from the L2 cache is hidden, guaranteeing an improved L1 cache hit rate.

Since the accumulating phase of the BRMerge has a typically streaming access pattern when merging the intermediate lists, BRMerge achieves the maximized memory bandwidth utilization and takes advantage of the hardware pre-fetching mechanism. In contrast, the hash-based and hashvec-based accumulation methods do not have the streaming access pattern since the hashing operation can direct the insertion to potentially any place in a hash table. Therefore, when the random memory accesses miss the L1 cache and load a cache line from the L2 cache, most memory bandwidth between L1 and L2 may be wasted. Moreover, the hash-based and hashvec-based accumulation methods cannot take advantage of the pre-fetching mechanism.

\subsubsection{Minimized Translation Lookaside Buffer (TLB) Cache Misses}
We discuss the TLB hit rate for two reasons: 1) both the row-wise and outer-product SpGEMM algorithms have to randomly access a large matrix (either the $B$ matrix for the row-wise SpGEMM or the $\hat{C}$ matrix for the outer-product SpGEMM~\cite{PB-SpGEMM}), which may cause high pressure for the TLB cache, and 2) the TLB cache hit rate is also critical for the performance of CPU programs. Some may argue that the outer-product SpGEMM can merge the partial matrices on the fly while generating them. However, this will require too many merging operations and data movement operations, which quickly cause more overheads than the gains, especially for the medium-to-large-size matrices which cannot fit in the cache~\cite{PB-SpGEMM}. Therefore, we do not consider that situation. 

We first explain the importance of the hit rate of L1-d TLB (i.e., the first level data TLB). Each time the CPU accesses a memory space, it generates a virtual address. The virtual address should be translated to a physical address (with page number) by consulting the L1-d TLB before finishing the querying to the L1 cache. If accessing the L1-d TLB misses, the querying to the L1 cache is hanged, and the CPU may consult the shared L2 TLB, which has a similar cost to that of accessing the L2 cache~\cite{intel-opt, DLL-TLB}. Therefore, the L1-d TLB miss also causes a similar cost to L2 cache access. Also, note that if accessing the L2 TLB misses again, the CPU should start a page table walk process, which involves much more expensive main memory accesses~\cite{DLL-TLB}.

In most cases, one TLB miss penalty can be alleviated by multiple following accesses to the same 4KiB page, which always hits the L1-d TLB. However, if each TLB miss is not followed with many accesses to the same 4KiB page, the penalty will of course not be alleviated, which is the case for the Heap-SpGEMM. We then shall discuss the possible TLB misses for the interested SpGEMM algorithm, how the proposed BRMerge accumulation method minimizes the TLB misses, and the inefficient TLB access issue in the Heap-SpGEMM.

We provide an upper-bound estimation of the possible TLB misses for the row-wise SpGEMM algorithm on Intel's recent Skylake server-class CPU. The Skylake CPU has a 64-entry private L1-d TLB per physical core backed by a 1536-entry shared L2 TLB per CPU~\cite{intel-opt}. The hyper-threading technology is often enabled to improve the performance of the multi-threaded programs~\cite{intel-opt}. When the hyper-threading technology is enabled, there will be two logical cores in each physical core, which share the L1-d TLB resources~\cite{intel-opt}. As a result, each logical core only has 32 entries of the L1-d TLB.

The upper-bound estimation assumes that the memory distance of different accessed rows in $B$ to compute one result row is more than the 4KiB page size. In other words, accessing each different row in $B$ occupies two TLB entries, where the "two" entries are occupied by the $col$ and $val$ arrays of a row. Therefore, when a logical core computes a result row that needs to access $32/2 = 16$ different rows in $B$, there may be capacity-caused L1-d TLB misses. Note that the number of the accessed rows of $B$ is the same as the $n_{nz}$ of the corresponding row in $A$. In the real situation, the $n_{nz}$ of rows in A can be much larger than 16, which can cause many TLB misses in the SpGEMM algorithm. 

Yet, the capacity-caused TLB misses are hard to be avoided in performing SpGEMM. However, minimizing the TLB misses can be easily achieved by ensuring that the TLB is only updated once when loading each row from the $B$ matrix. To achieve this, in many cases, we only need to make sure each row of the $B$ matrix is loaded consecutively in a short duration. The proposed BRMerge accumulation method meets this requirement since it accesses all the required rows in $B$ consecutively and stores the multiplied intermediate lists in a local ping-pong buffer in the multiplying phase.

In contrast, the Heap-SpGEMM does not meet the above requirement. It maintains the accesses to all the required rows of $B$ during the its accumulating phase. Therefore, the corresponding TLB entries for accessing each row of $B$ may be inserted into and evicted out of the TLB multiple times, suffering not-alleviated TLB misses penalties when there are capacity-caused TLB misses.

\subsubsection{Reasonably High L1/L2 Cache Hit Rates}
When accumulating the multiple intermediate lists in the accumulation method, the data may be processed multiple times in the L1/L2 cache. Therefore, the L1/L2 cache hit rates are important for the performance of the accumulation method. 

Since BRMerge is a row-wise accumulation method, it only processes the intermediate lists for one result row in each iteration, which is usually small and comparable to the L1/L2 cache size. Moreover, after each round of merging, the required memory size of the intermediate lists is reduced due to the merging of duplicate indices. Therefore, the expected cache hit rates of the L1/L2 caches increases after each round of merging. Furthermore, since BRMerge has a typical streaming access pattern during the accumulating phase, the hardware pre-fetching mechanism further increases the L1/L2 cache hit rates. As a result, the L1/L2 cache hit rates of BRMerge during the accumulating phase is reasonably high.

In summary, the architectural benefits of the proposed BRMerge accumulation method are the streaming access patterns, minimized TLB misses, and reasonably high L1/L2 cache hit rates, which result in both low access latency and high bandwidth utilization when performing SpGEMM on the multi-core CPU architectures.

\subsection{BRM\lowercase{erge}-Upper and BRM\lowercase{erge}-Precise}\label{sec:spgemm}
To implement a high-performance SpGEMM algorithm, the allocation and the load balance methods are also important~\cite{yusuke, nsparse}. Since the upper-bound and precise allocation methods are commonly used and yields high performance~\cite{bhsparse, yusuke}, we use the two allocation methods and implements two SpGEGMM libraries named BRMerge-Upper and BRMerge-Precise. The two SpGEMM libraries are based on the BRMerge accumulation method.

The load balance of the two SpGEMM algorithms is the same as the previous work~\cite{yusuke}, which is reported to have high performance. First, the $n_{prod}$ of computing each output row is counted. Then the work is statically divided into $p$ groups in a per-row granularity, where $p$ is the number of CPU threads. The division policy is to keep the same total $n_{prod}$ in each group of work. Each group of work is computed by one CPU thread.

\subsubsection{BRM\lowercase{erge}-Upper}\label{sec:upper}
\begin{figure}[h]
\centering
\includegraphics[width=0.45\textwidth]{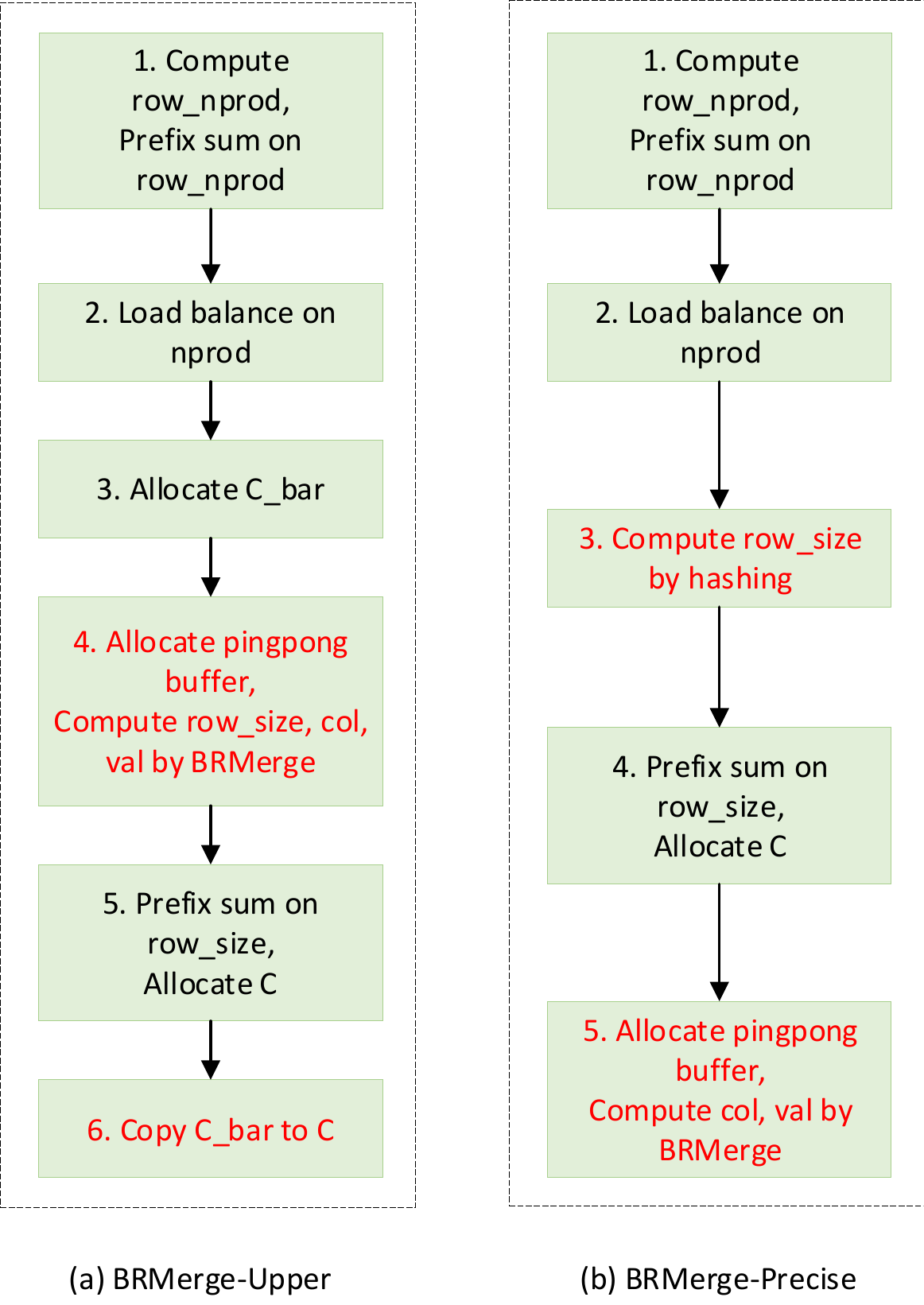}
\caption{Computation steps of BRMerge-Upper (Left) and BRMerge-Precise (Right)}
\label{fig:spgemm}
\end{figure}

Fig.~\ref{fig:spgemm}~(a) shows the computation steps of the BRMerge-Upper SpGEMM algorithm. Step~1 computes the \emph{row\_nprod} of each output row and performs prefix sum on the \emph{row\_nprod} for two purposes: 1) load balance (step~2) and 2) allocating the temporary memory space for the $C$ matrix (step~3). The temporary memory space for the $C$ matrix is denoted as $C\_bar$. Step~4 allocates the ping-pong buffers and computes the \emph{row\_size}, $col$, and $val$ arrays of the $C$ matrix by the BRMerge accumulation method. The \emph{row\_size} represents each row's number of nonzero elements. Step~5 performs prefix sum on the \emph{row\_size} to obtain the $rpt$ array and total $n_{nz}$ of the $C$ matrix at the same time. The $col$ and $val$ arrays of $C$ are allocated according to the total $n_{nz}$ of $C$. Step~6 copies the $C\_bar$ matrix to the standard $C$ matrix which conforms to the CSR storage format.

Step~3 allocates the $C\_bar$ matrix parallelly, which means each CPU thread only allocates the part of $C\_bar$ for its private use. The load balance of step~4 and step~6 is based on the $n_{prod}$ information, which means each CPU thread computes approximately the same number of intermediate products. Whereas, the load balance of step~1 and step~5 is based on the number of rows of $C$, which means each CPU thread computes approximately the same number of rows. Step~3 is computed in a single CPU thread since its computation complexity is very low (i.e., $O(p)$, where $p$ is the number of CPU threads).

\subsubsection{BRM\lowercase{erge}-Precise}\label{sec:precise}
Fig.~\ref{fig:spgemm}~(b) shows the computation steps of the BRMerge-Precise SpGEMM algorithm. Step~1 computes the \emph{row\_nprod} of each output row and performs prefix sum on the \emph{row\_nprod} for the load balance in step~2. Step~3 computes the \emph{row\_size} of the $C$ matrix by the hashing method described in the previous work~\cite{yusuke}. Step~4 performs the prefix sum on the computed \emph{row\_size} to obtain the $rpt$ array and the total $n_{nz}$ of $C$. Then the $col$ and $val$ arrays of $C$ are allocated according to the total $n_{nz}$ of $C$. Step~5 allocates the ping-pong buffers and computes the $col$ and $val$ arrays by the BRMerge accumulation method. Each computed row is directly written into the $col$ and $val$ arrays, which conform to the CSR storage format.

The load balance of step~3 and step~5 is based on the $n_{prod}$ information. Whereas the load balance of step~1 and step~4 is based on the number of rows of the $C$ matrix.

\section{Performance Evaluation}\label{sec:evaluate}
\subsection{Evaluation Setup}\label{sec:env}

We compare the proposed SpGEMM libraries BRMerge-Upper and BRMerge-Precise to Heap-SpGEMM~\cite{yusuke}, Hash-SpGEMM~\cite{yusuke}, Hashvec-SpGEMM~\cite{yusuke}, PB-SpGEMM~\cite{PB-SpGEMM}, and MKL-SpGEMM~\cite{mkl-doc}. The evaluation is based on the FLOPS performance of the matrix square benchmarks~\cite{bhsparse, nsparse, speck}, which is twice the number of the intermediate products divided by the execution time. The execution time is obtained by averaging the execution time of ten runs after one warm-up run. In all benchmarks, we measure the execution time of the approaches in double precision.

\begin{table}[h]
\caption{Detailed Information about the two CPU servers and the software environment.}
\fontsize{7}{10.5}\selectfont
\label{tab:env}
\centering
\begin{tabular}{|l|l|l|}
	\hline 
	\textbf{Server name} & Platinum & Gold \\
	\hline
	CPU name & Intel Xeon Platinum 8163 & Intel Xeon Gold 6254 \\
	\hline
	\#CPUs & 2 & 4 \\
	\hline
	\#Cores/CPU & 24 & 18 \\
	\hline
	\#Threads/core & 2 & 1 \\
	\hline
	\#Threads & 96 & 72\\
	\hline
	\#Threads (\textbf{used}) &  \textbf{80} &  \textbf{58}\\
	\hline
	Clock frequency & 2.7GHz & 3.1GHz \\
	\hline
	L1-d cache & 32KiB/core & 32KiB/core \\
	\hline
	L1-i cache & 32KiB/core & 32KiB/core \\
	\hline
	L2 cache & 1024KiB/core & 1024KiB/core \\
	\hline
	L3 cache & 33MiB/CPU & 24.75MiB/CPU \\
	\hline
	Main memory type & DDR4  & DDR4\\
	\hline
	Main memory capacity & 755.1GiB & 4031.2GiB \\
	\hline
	Linux kernel & \multicolumn{2}{l|}{3.10.0} \\
	\hline
	Compiler & \multicolumn{2}{l|}{Intel icc (ICC) 2021.6.0 20220226} \\
	\hline
	Compiler option & \multicolumn{2}{l|}{-O2 -fopenmp -std=c++17} \\ 
	\hline
\end{tabular}

\end{table}

\begin{table*}[ht]
\centering
\caption{Detailed information about the 26 sparse matrices.}
\label{tab:matrix}
\begin{tabular}{|c|c|c|c|c|c|c|c|c|}
	\hline
	\textbf{Id} & \textbf{Name} & \textbf{Rows} & \textbf{Nnz} & \textbf{Nnz/row} & \textbf{Max nnz/row} & \textbf{Nprod of $A^2$} & \textbf{Nnz of $A^2$} & \textbf{Compression ratio of $A^2$} \\
	\hline
	1 & m133-b3 & \num{200200} & \num{800800} & 4.0 & \num{4} & \num{3203200} & \num{3182751} & 1.01 \\
	\hline
	2 & mac\_econ\_fwd500 & \num{206500} & \num{1273389} & 6.2 & \num{44} & \num{7556897} & \num{6704899} & 1.13 \\
	\hline
	3 & patents\_main & \num{240547} & \num{560943} & 2.3 & \num{206} & \num{2604790} & \num{2281308} & 1.14 \\
	\hline
	4 & webbase-1M & \num{1000005} & \num{3105536} & 3.1 & \num{4700} & \num{69524195} & \num{51111996} & 1.36 \\
	\hline
	5 & mc2depi & \num{525825} & \num{2100225} & 4.0 & \num{4} & \num{8391680} & \num{5245952} & 1.60 \\
	\hline
	6 & scircuit & \num{170998} & \num{958936} & 5.6 & \num{353} & \num{8676313} & \num{5222525} & 1.66 \\
	\hline
	7 & delaunay\_n24 & \num{16777216} & \num{100663202} & 6.0 & \num{26} & \num{633914372} & \num{347322258} & 1.83 \\
	\hline
	8 & mario002 & \num{389874} & \num{2101242} & 5.4 & \num{7} & \num{12829364} & \num{6449598} & 1.99 \\
	\hline
	9 & cage15 & \num{5154859} & \num{99199551} & 19.2 & \num{47} & \num{2078631615} & \num{929023247} & 2.24 \\
	\hline
	10 & cage12 & \num{130228} & \num{2032536} & 15.6 & \num{33} & \num{34610826} & \num{15231874} & 2.27 \\
	\hline
	11 & majorbasis & \num{160000} & \num{1750416} & 10.9 & \num{11} & \num{19178064} & \num{8243392} & 2.33 \\
	\hline
	12 & wb-edu & \num{9845725} & \num{57156537} & 5.8 & \num{3841} & \num{1559579990} & \num{630077764} & 2.48 \\
	\hline
	13 & offshore & \num{259789} & \num{4242673} & 16.3 & \num{31} & \num{71342515} & \num{23356245} & 3.05 \\
	\hline
	14 & 2cubes\_sphere & \num{101492} & \num{1647264} & 16.2 & \num{31} & \num{27450606} & \num{8974526} & 3.06 \\
	\hline
	15 & poisson3Da & \num{13514} & \num{352762} & 26.1 & \num{110} & \num{11768678} & \num{2957530} & 3.98 \\
	\hline
	16 & filter3D & \num{106437} & \num{2707179} & 25.4 & \num{112} & \num{85957185} & \num{20161619} & 4.26 \\
	\hline
	17 & cop20k\_A & \num{121192} & \num{2624331} & 21.7 & \num{81} & \num{79883385} & \num{18705069} & 4.27 \\
	\hline
	18 & mono\_500Hz & \num{169410} & \num{5036288} & 29.7 & \num{719} & \num{204030968} & \num{41377964} & 4.93 \\
	\hline
	19 & conf5\_4-8x8-05 & \num{49152} & \num{1916928} & 39.0 & \num{39} & \num{74760192} & \num{10911744} & 6.85 \\
	\hline
	20 & cant & \num{62451} & \num{4007383} & 64.2 & \num{78} & \num{269486473} & \num{17440029} & 15.45 \\
	\hline
	21 & hood & \num{220542} & \num{10768436} & 48.8 & \num{77} & \num{562028138} & \num{34242180} & 16.41 \\
	\hline
	22 & consph & \num{83334} & \num{6010480} & 72.1 & \num{81} & \num{463845030} & \num{26539736} & 17.48 \\
	\hline
	23 & shipsec1 & \num{140874} & \num{7813404} & 55.5 & \num{102} & \num{450639288} & \num{24086412} & 18.71 \\
	\hline
	24 & pwtk & \num{217918} & \num{11634424} & 53.4 & \num{180} & \num{626054402} & \num{32772236} & 19.10 \\
	\hline
	25 & rma10 & \num{46835} & \num{2374001} & 50.7 & \num{145} & \num{156480259} & \num{7900917} & 19.81 \\
	\hline
	26 & pdb1HYS & \num{36417} & \num{4344765} & 119.3 & \num{204} & \num{555322659} & \num{19594581} & 28.34 \\
	\hline
	
\end{tabular}
\end{table*}

We evaluate all the approaches on two CPU servers with different CPUs. All the evaluated approaches are compiled and executed with the same hardware and software environment on each CPU server. The detailed information about the two CPU servers and the software environment is summarized in Table~\ref{tab:env}. 

We select 26 square matrices from the SuiteSparse Matrix Collection~\cite{suitesparse} for the evaluation. The selected matrices are commonly used for the performance evaluation of SpGEMM on CPUs~\cite{yusuke, PB-SpGEMM} and GPUs~\cite{bhsparse, nsparse, speck}. The detailed information of these matrices is summarized in Table~\ref{tab:matrix}. Note that the 26 matrices in Table~\ref{tab:matrix} are sorted in an ascending order by their compression ratio.

The MKL-SpGEMM is implemented based on the C version of Intel's oneAPI Math Kernel Library (oneMKL)~\cite{mkl-doc}. The key routine in the MKL-SpGEMM is the \emph{mkl\_sparse\_spmm} routine. The parallelism in all the approaches including MKL-SpGEMM is based on the OpenMP framework~\cite{openmp}. Therefore, we use \emph{omp\_set\_num\_threads} to set the used number of threads ($\#threads$). All the evaluations are set with the same $\#threads$ on each CPU server~(Table~\ref{tab:env}).

During the performance evaluation, we observe that when the used $\#threads$ is set as the maximum $\#threads$ on a CPU server, which is 96 on the Platinum server and 72 on the Gold server, even the average execution time can vary widely in different runs. The possible reasons for this might be that the occasional operating system services may seize several CPU threads for execution and evict the executing SpGEMM tasks. Based on our experiments, finally, we set the $\#threads$ as 80 on the Platinum server and 58 on the Gold server for all the evaluations and observed a relatively stable average execution time among different runs. 

Another performance note is that the \emph{scalable\_malloc} and \emph{scalable\_free} in Intel's oneAPI Thread Building Block (oneTBB)~\cite{tbb-doc} are used in the state-of-the-art SpGEMM libraries including Heap-SpGEMM, Hash-SpGEMM, Hashvec-SpGEMM, and PB-SpGEMM. And the \emph{scalable\_malloc} and \emph{scalable\_free} yield better performance for the parallel memory allocation and deallocation~\cite{yusuke}. Therefore, we also use the \emph{scalable\_malloc} and \emph{scalable\_free} in our SpGEMM implementations. 

\subsection{Results and Discussions}\label{sec:result}

\subsubsection{Performance of S\lowercase{p}GEMM on the Platinum Server}

\begin{figure*}[ht]
\centering
\includegraphics[width=0.95\textwidth]{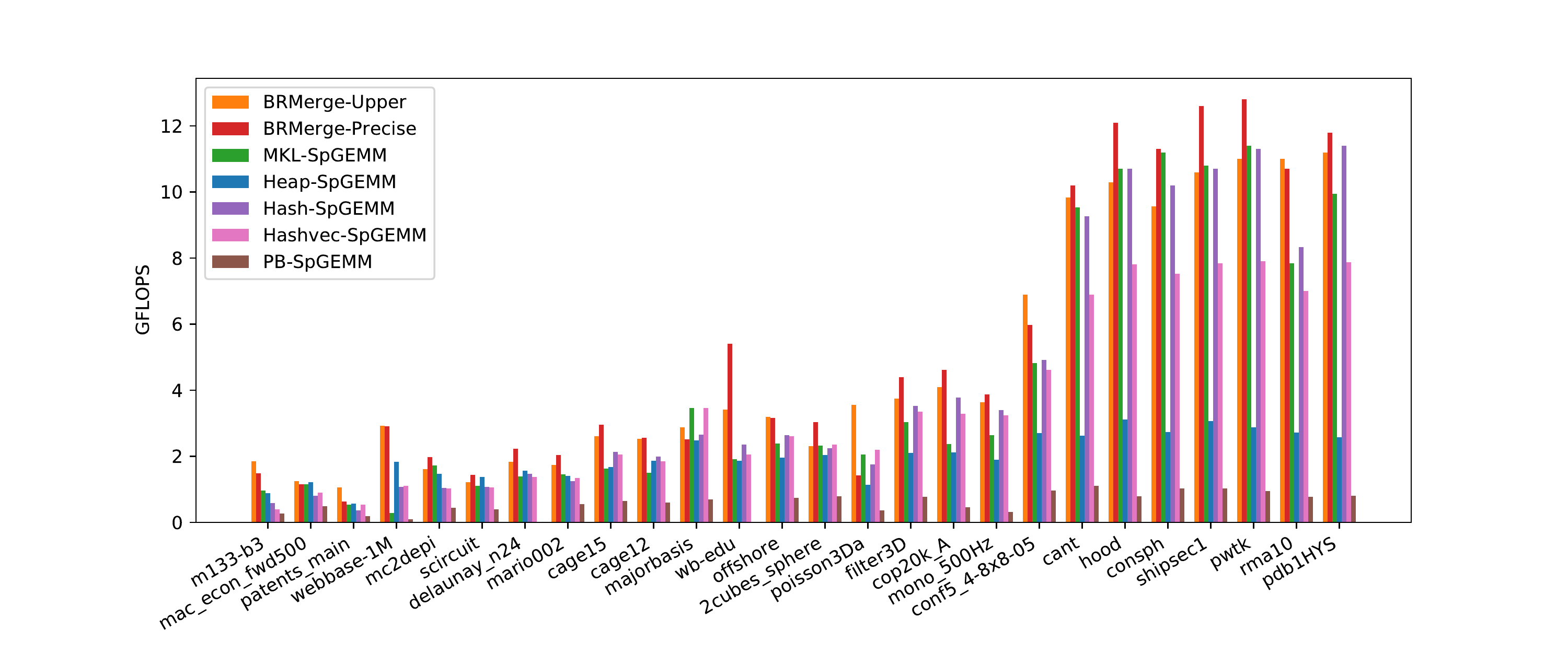}
\caption{Performance of SpGEMM on the Platinum server}
\label{fig:perf-cs08}
\includegraphics[width=0.95\textwidth]{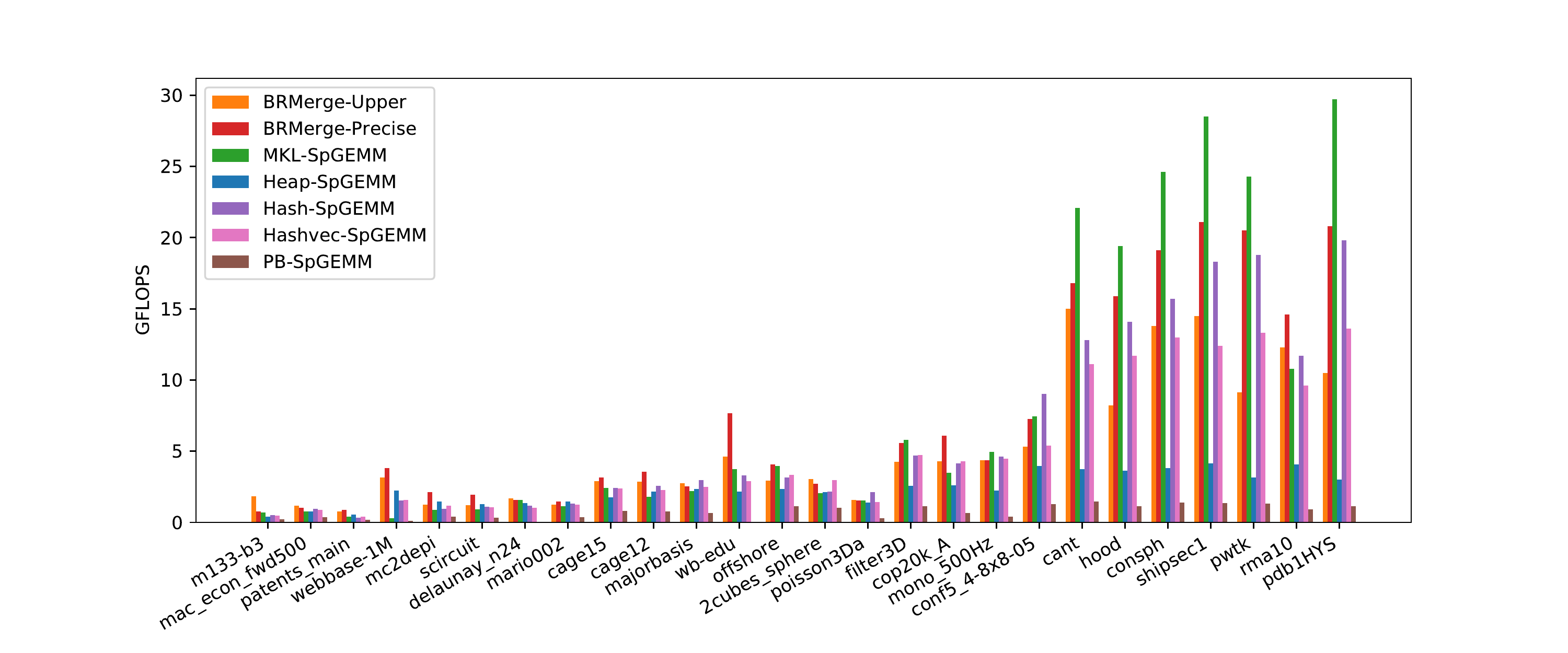}
\caption{Performance of SpGEMM on the Gold server}
\label{fig:perf-cs11}
\end{figure*}

Fig.~\ref{fig:perf-cs08} shows the overall performance of the evaluated SpGEMM libraries on the Platinum server. The results show that the BRMerge-Precise outperforms other libraries on most benchmarks. The average performance speedups of BRMerge-Precise over BRMerge-Upper, Heap-SpGEMM, Hash-SpGEMM, Hashvec-SpGEMM, MKL-SpGEMM, and PB-SpGEMM are $1.05\times$, $2.29\times$, $1.42\times$, $1.56\times$, $1.68\times$, $8.46\times$, respectively. The best-performing and the second best-performing SpGEMM algorithms are the proposed BRMerge-Precise and BRMerge-Upper. Moreover, the BRMerge-Precise achieves an average $1.42\times$ performance speedup over the state-of-the-art best-performing SpGEMM library, which is the Hash-SpGEMM.

The PB-SpGEMM suffers exceptionally lower performance on \textbf{all} the tested benchmarks, including benchmarks with low compression ratios. We attribute the exceptionally low performance of PB-SpGEMM to the outer-product dataflow, which suffers worse memory access efficiency to the $\hat{C}$ matrix compared to the memory access efficiency to the $B$ matrix for the row-wise dataflow. The previous work also reported that the outer-product dataflow has significant drawbacks compared to the row-wise dataflow~\cite{matraptor, gamma}.

We compare the performance of the BRMerge-Upper with the Heap-SpGEMM since the two SpGEMM libraries implement the same allocation and load balance methods. For the benchmarks with low compression ratios and low $n_{nz}$ per row of the $A$ matrix (i.e., the matrix with ids ranges from 1 to 8, except for the webbase-1M matrix), the performance of the Heap-SpGEMM is comparable to the proposed BRMerge-Upper SpGEMM library. The reason is that the two main drawbacks of the Heap-SpGEMM are avoided in these benchmarks, which are 1) the high computation complexity on benchmarks with high compression ratios~\cite{yusuke}, and 2) the high TLB cache miss rate due to accessing the rows of the $B$ matrix in a longer duration (Section~\ref{sec:memory}). For other evaluated benchmarks, BRMerge-Upper significantly outperforms Heap-SpGEMM despite the two SpGEMM libraries implementing the same allocation and load balance methods. For the evaluated benchmarks with the matrix ids above 20, BRMerge-Upper even achieves more than $3\times$ better peroformance than the Heap-SpGEMM. We attribute this large performance gap to the architectural benefits of the proposed accumulation method and the high TLB cache miss rate of the Heap-SpGEMM (Section~\ref{sec:memory}).

We compare the performance of the BRMerge-Precise with the Hash-SpGEMM since the two SpGEMM libraries implement the same allocation and load balance methods. Moreover, the symbolic phase of the BRMerge-Precise also uses the hash-based method to count the $n_{nz}$ per row~\cite{yusuke} which is the same as the Hash-SpGEMM. The Hash-SpGEMM is reported to have better performance than other SpGEMM libraries on benchmarks with high compression ratios~\cite{yusuke}. The reason is that, when combined with the precise allocation method, the $n_{nz}$ per output row is known before the numeric phase; therefore, the accessed hash table size can be as small as the $n_{nz}$ per output row. 

Nevertheless, for benchmarks with relatively high compression ratios (i.e., the matrices with ids ranging from 15 to 26), the BRMerge-Precise outperforms the Hash-SpGEMM. We attribute this to three reasons. First, the BRMerge accumulation method has streaming memory access patterns, which can achieve maximized bandwidth utilization and increase the L1 cache hit rate by taking advantage of the hardware pre-fetching mechanism. Whereas, the hash-based accumulation method does not have the architectural benefits mentioned above. Second, the BRMerge accumulation method has reasonably high L1/L2 cache hit rates as a row-wise accumulation method (section ~\ref{sec:memory}). Third, the result list in the BRMerge accumulation method is natively sorted. However, the hash-based accumulation method has to perform the additional extracting and sorting operations after the hashing operation. 

As for the benchmarks with low compression ratios, the BRMerge-Precise significantly outperforms the Hash-SpGEMM. Note that in our evaluation, the Hashvec-SpGEMM seldom outperforms the Hash-SpGEMM in the evaluated benchmarks.

\subsubsection{Performance of S\lowercase{p}GEMM on the Gold Server}
Fig.~\ref{fig:perf-cs11} shows the overall performance of the evaluated SpGEMM libraries on the Gold server. The BRMerge-Precise achieves on average $1.29\times$, $2.62\times$, $1.39\times$, $1.50\times$, $1.73\times$, and $9.37\times$ performance speedup over the BRMerge-Upper, Heap-SpGEMM, Hash-SpGEMM, Hashvec-SpGEMM, MKL-SpGEMM, and PB-SpGEMM, respectively. The best-performing and second-best-performing SpGEMM libraries are still the proposed BRMerge-Precise and BRMerge-Upper. Moreover, the state-of-the-art best-performing SpGEMM library is the Hash-SpGEMM.

The performance trends of all the evaluated SpGEMM libraries except for the MKL-SpGEMM on the Gold server are similar to those on the Platinum Server. The MKL-SpGEMM outperforms the proposed BRMerge-Precise on benchmarks with high compression ratios (i.e., the matrices with ids ranging from 18 to 26). In contrast, the proposed BRMerge-Precise outperforms the MKL-SpGEMM on benchmarks with compression ratios lower than 4. Note that on the highly irregular webbase-1M benchmark, the BRMerge-Precise significantly outperforms the MKL-SpGEMM on both the Platinum server and the Gold server. On average, the BRMerge-Precise outperforms the MKL-SpGEMM by $1.73\times$. However, we could not explain the reasons for this performance difference since MKL-SpGEMM's detailed implementation is unavailable.


\section{Conclusion}\label{sec:conclude}
In this paper, we present a novel binary-row-merging-based accumulation method, named BRMerge, to optimize the performance of SpGEMM on multi-core CPU architectures. We also analyze the architectural benefits of the proposed accumulation method, which are the streaming access patterns, minimized TLB cache miss rate, and reasonably high L1/L2 cache hit rates. As a result, the proposed accumulation method can achieve both low access latency and high bandwidth utilization when performing SpGEMM on multi-core CPU architectures. Based on the proposed accumulation method, we further propose two parallel SpGEMM libraries named BRMerge-Precise and BRMerge-Upper based on different allocation methods. The proposed SpGEMM libraries significantly outperform the state-of-the-art SpGEMM libraries on 26 commonly used matrices with two CPU servers.  Specifically, our best-performing SpGEMM library (i.e., BRMerge-Precise) achieves on average $1.42\times$ and $1.39\times$ performance speedups compared to the state-of-the-art best-performing SpGEMM library (i.e., Hash-SpGEMM~\cite{yusuke}) on the Intel Xeon Platinum 8163 CPU and the Intel Xeon Gold 6254 CPU, respectively.





\EOD
\end{document}